# Nonlinear dielectric response at the excess wing of glass-forming liquids

Th. Bauer, P. Lunkenheimer,[*] S. Kastner, and A. Loidl

*Experimental Physics V, Center for Electronic Correlations and Magnetism, University of Augsburg, 86135 Augsburg, Germany*

We present nonlinear dielectric measurements of glass-forming glycerol and propylene carbonate applying electrical fields up to 671 kV/cm. The measurements extend to sufficiently high frequencies to allow for the investigation of the nonlinear behavior in the regime of the so-far mysterious excess wing, showing up in the loss spectra of many glass formers as a second power law at high frequencies. Surprisingly, we find a complete lack of nonlinear behavior in the excess wing, in marked contrast to the $\alpha$ relaxation where, in agreement with previous reports, a strong increase of dielectric constant and loss is found.

PACS numbers: 77.22.Gm, 64.70.pm, 77.22.Ch

The measurement of susceptibility spectra is an invaluable method in the investigation of the still poorly understood slowing down of molecular motion at the glass transition. There, usually the linear response of a glass former to an applied external field is detected. A prominent example is the permittivity measured by dielectric spectroscopy [1,2]. However, stimulated by great successes in the study of spin and orientational glasses [3,4,5], in recent years the investigation of the nonlinear response of glass-forming matter is attracting increasing interest (see, e.g., [6,7,8,9,10,11]). Here again dielectric spectroscopy has played an important role [8,10,12,13,14,15,16,17,18]. A notable example is the experimental proof of heterogeneity by dielectric hole-burning (DHB) [19]. Furthermore, critical information on the correlation length scales was obtained by measurements of the higher order susceptibility $\chi_3$ [10,12,18]. Moreover, nonlinear contributions to the permittivity were found to be directly related to the heterogeneous distribution of relaxation times [8,15]. It should be noted that all these investigations were essentially confined to the $\alpha$-relaxation regime. The $\alpha$ relaxation corresponds to the structural rearrangement of the molecules that governs, e.g., viscous flow. However, in recent years the attention of many scientists has switched to additional, faster processes, which are believed to be of high relevance for the understanding of the glass transition (see, e.g., [1,2,20,21,22,23,24]). Until now, not much is known about their nonlinear properties, except for results from DHB experiments (e.g., [25,26]). The reason may be the small nonlinear susceptibilities associated to these processes, as even the corresponding linear susceptibilities are much smaller than those of the $\alpha$ relaxation.

In the present work, we provide results on the nonlinear behavior of the so-called excess wing (EW), found in a variety of partly very different glass formers like glycerol [1,27], ionic melts [28], or even metallic glasses [29]. In spectra of the dielectric loss $\varepsilon''$, it shows up as a second power law at the high-frequency flank of the peak caused by the $\alpha$ relaxation. The EW was shown to be due to a secondary relaxation, faster than the $\alpha$ relaxation [21,22,24], whose origin is unclear up to now. One may speculate [21,22,30] that it is a manifestation of the Johari-Goldstein $\beta$-relaxation, revealed in various glass formers by a loss peak arising at a frequency beyond that of the $\alpha$ peak [31,32,33]. However, other interpretations were also discussed [23,32]. Unfortunately, the origin of the Johari-Goldstein relaxation also is unclear until now. Interestingly, DHB experiments revealed different behavior in the regimes of the EW and the $\alpha$ relaxation of glass-forming glycerol concerning the persistence time of the burned holes [26]. Moreover, physical aging measurements (which can be regarded as a special type of nonlinear experiments), performed in the EW region of glycerol [34], could be explained without invoking the typical heterogeneous dynamics of glass-forming matter, in marked contrast to the $\alpha$-relaxation regime [35]. Thus, a thorough investigation of the nonlinear dielectric properties of the EW seems of vital importance and may help clarifying its origin.

Nonlinear dielectric spectroscopy can be performed in different ways, e.g., by detecting the difference $\Delta\varepsilon$ of measurements with low and high ac field [8,15] or by the determination of higher harmonics of the permittivity [10,12,18]. It should be noted that these measurements provide distinctly different information [36]. In the present work, we follow the first approach. In the seminal paper by Richert and Weinstein [8], this method was applied to glycerol. A significant increase of $\varepsilon''$ was found at the high-frequency flank of the $\alpha$ peak while $\Delta\varepsilon''$ was nearly zero at low frequencies. These results were fully consistent with a model assuming dynamical heterogeneities with closely correlated dielectric and thermal relaxation times. Its basic ideas were also used to explain DHB [19] and it also is of high relevance to understand the microwave heating of liquids [37]. In the present work, we provide data on the high-field behavior of glycerol and propylene carbonate (PC) covering higher fields and a frequency and temperature range that extends well into the region of the EW.

The measurements were performed using a frequency-response analyzer in combination with a high voltage booster "HVB 300", both from Novocontrol Technologies, enabling measurements with peak voltages up to 150 V at frequencies up to about 100 kHz. The sample material was put between two lapped and highly polished steel plates, forming a thin film with thickness of 2.2 μm (glycerol) and 1.1 μm (PC). No spacer materials were used, which significantly reduced the probability for electrical breakthroughs. For a verification of the obtained results, additional measurements with glass-fiber spacers of 30 μm diameter were carried out, using a high voltage booster "HVB 4000" and a ferroelectric analyzer (aixACCT TF2000), reaching voltages up to 1.1 kV. Similar to the procedure reported in [8], at each frequency we have performed successive high- and low-field measurements, separated by a waiting time. At low frequencies, fewer high-voltage cycles than in [8] were applied (typically 2-5 [38]) to fully exclude trivial heating effects while at higher frequencies the cycle number was larger and determined by the processing speed of the experimental setup. Using the mechanical modulus of glycerol, the sample contraction due to the attracting force between the capacitor plates was estimated to be less than 0.5% [38].

Figure 1 shows the dielectric constant $\varepsilon'$ and loss $\varepsilon''$ of glycerol as obtained for fields of 14 and 671 kV/cm at four selected temperatures. The lines represent previously published spectra, measured with 0.2 kV/cm [39] (at 186 K, no data from [39] are available). They reasonably agree with the spectra at 14 kV/cm [40]. This demonstrates that at 14 kV/cm, the material still is in the linear regime and that there is negligible "phonon-bath heating" [8] by the preceding high-voltage cycle. In contrast, the high-field data exhibit marked deviations from the linear results, predominantly showing up in the region of the high-frequency flanks of the α peaks and reaching values of up to 50% for $\varepsilon''$ and 40% for $\varepsilon'$. In contrast, at the peak frequency $\nu_p$ and below, no significant field dependence is detected. These findings in the α-peak regime are in good qualitative accord with those reported by Richert and Weinstein where in $\varepsilon''$ at 213 K an increase of 8.6% was found for a lower field of 282 kV/cm [8]. Applying the same field as in [8] also leads to a good quantitative agreement (not shown).

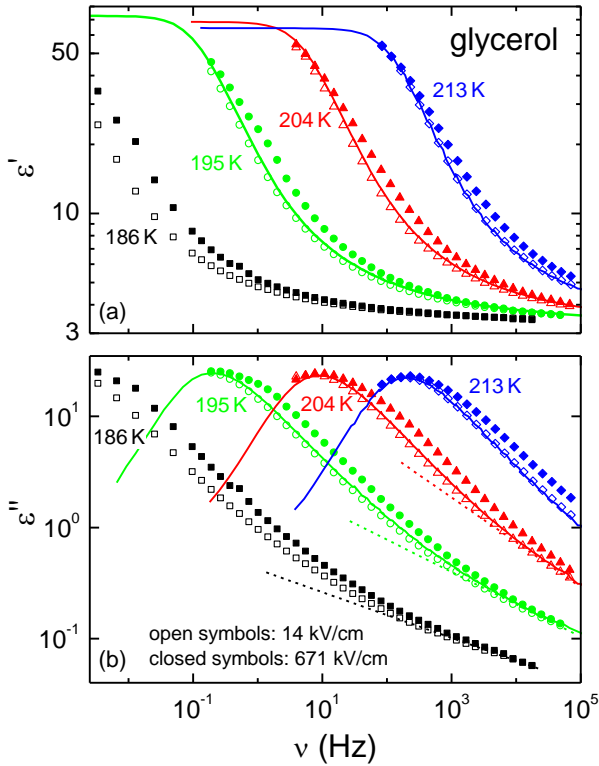

FIG. 1 (color online). $\varepsilon'$ (a) and $\varepsilon''$ (b) of glycerol measured at fields of 14 kV/cm (open symbols) and 671 kV/cm (closed symbols) shown for four temperatures. The lines correspond to results published in [39], measured with 0.2 kV/cm [40].

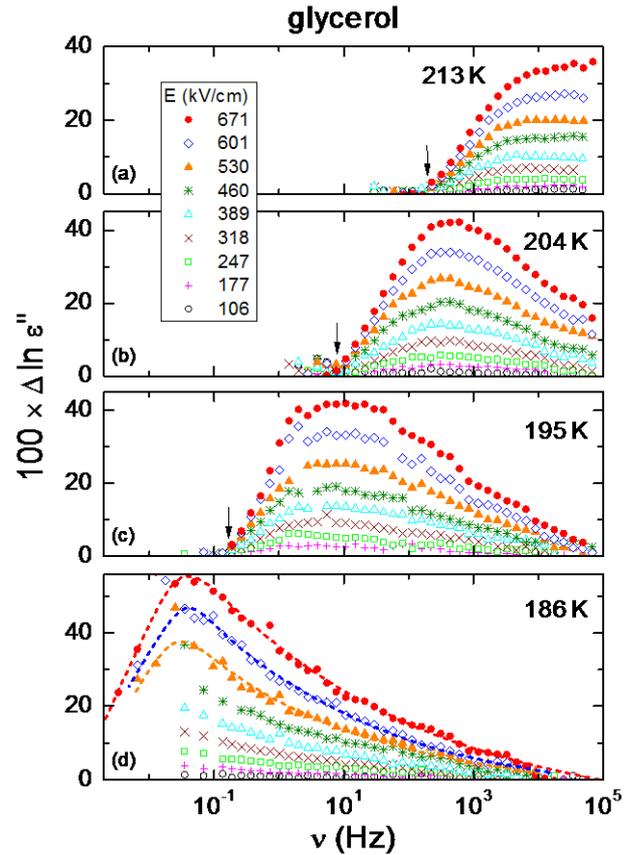

FIG. 2 (color online). Difference of the logarithm of the loss spectra of glycerol, measured with various high fields and with a low field of 14 kV/cm. In frames (a) - (d), results for four different temperatures are shown. The arrows indicate the α-peak positions (cf. Fig. 1). The lines in (d) are shown to guide the eyes.

In contrast to those reported in [8], the present spectra also extend into the EW region, which for the three lowest temperatures is indicated by the dashed lines in Fig. 1(b). Here the high- and low-field spectra approach each other and become practically identical. The same behavior is found in $\varepsilon'$. For a quantitative assessment, in Fig. 2 the difference $\Delta \ln \varepsilon''(E) = \ln \varepsilon''(E) - \ln \varepsilon''(14\ \mathrm{kV/cm})$ is shown for a variety of fields $E$. Similar to the results reported in [8], $\Delta \ln \varepsilon''$ is zero or even slightly negative for frequencies below the peak frequency (indicated by the arrows in Fig. 2) [38]. For $\nu > \nu_p$, the curves strongly increase and for 213 K [Fig. 2(a)] they saturate at the highest investigated frequencies. Despite some scatter of the data, such a saturation could already be suspected from the findings in [8] for a field of 282 kV/cm and it also is in good accord with the model predictions of that work. However, at 213 K the spectra are completely dominated by the $\alpha$ relaxation (Fig. 1). For lower temperatures, the EW shifts into the frequency window and in $\Delta \ln \varepsilon''$ a clear peak shows up that shifts to lower frequencies with decreasing temperature [Figs. 2(b) - (d)]. Most remarkably, at low temperatures and high frequencies, $\Delta \ln \varepsilon''$ finally reaches zero, even for the highest fields investigated [Figs. 2(c) and (d)]. This finding implies that deep in the EW regime of glycerol, no field dependence of $\varepsilon''$ exists. The same was found in measurements using spacers and two different devices, as noted in the experimental part.

To check if this unexpected finding is a general behavior of glass formers with EW, additional measurements of glass-forming PC were performed (Fig. 3). Again, a strong field-induced increase of $\varepsilon''$ is found at the high-frequency flank of the $\alpha$ peak while there is no significant field dependence at its low-frequency flank. Moreover, just as for glycerol, deep in the region of the EW, nicely revealed at 158 K, the field dependence vanishes. As shown in the inset of Fig. 3, $\Delta \ln \varepsilon''$ exhibits qualitatively similar behavior as for glycerol: a peak shows up, shifting to lower frequencies with decreasing temperature [38]. As seen for 158 K, its right flank approaches zero for high frequencies. The onset of a peak in $\Delta \ln \varepsilon''$ was also found in ref. [14] for PC at 166 K. It was explained within the model promoted in [8] by taking into account the smaller slope $d \ln \varepsilon'' / d \ln \nu$ in the EW region [14]. However, within this framework the peak in $\Delta \ln \varepsilon''$ should be followed by a constant plateau towards the highest frequencies. The experimental fact that $\Delta \ln \varepsilon''(\nu)$ approaches zero at high frequencies, revealed in the present work, cannot be explained in this way [38].

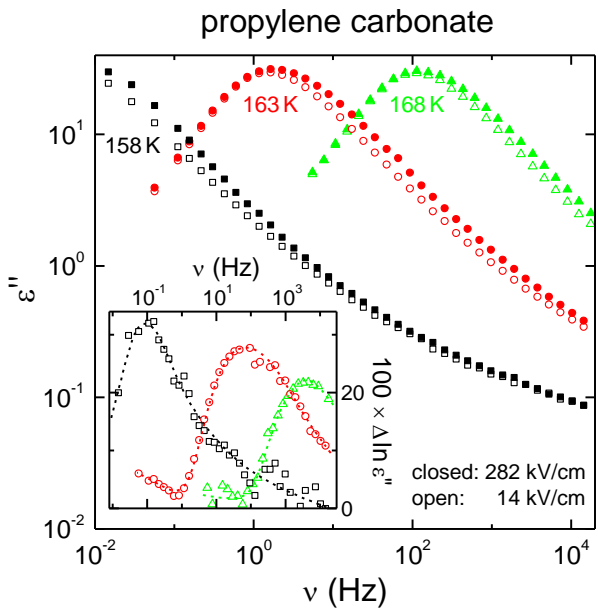

FIG. 3 (color online). $\varepsilon''$ of PC, measured at fields of 14 kV/cm (open symbols) and 282 kV/cm (closed symbols) shown for three temperatures. In the inset, the difference of the logarithm of the loss spectra shown in the main frame is plotted. The lines are guides for the eyes.

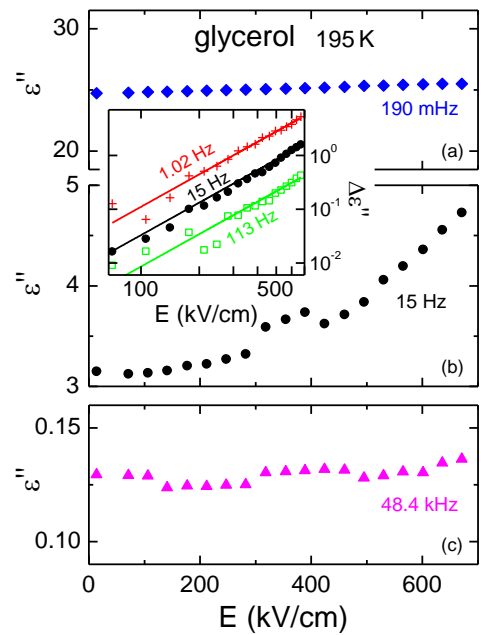

FIG. 4 (color online). Field dependence of $\varepsilon''$ of glycerol at 195 K, measured close to $\nu_p$ (a), at the high-frequency flank of the $\alpha$ peak (b), and in the EW region (c). The inset shows the difference of high- and low-field measurement for three frequencies located at the right flank of the $\alpha$ peak (double-logarithmic representation). The lines are linear fits with slope 2, corresponding to $\Delta \varepsilon'' \propto E^2$.

As an example of the field dependence of $\varepsilon''$, Fig. 4 shows the behavior for glycerol at 195 K. Close to the peak frequency (a), only very weak field dependence is found. In contrast, at the high-frequency flank of the $\alpha$ peak, a strong increase of $\varepsilon''(E)$ is observed [Fig. 4(b)]. These results agree with the findings in [8]. In the inset, $\Delta\varepsilon''(E) = \varepsilon''(E) - \varepsilon''(14\text{ kV/cm})$ is shown in a double-logarithmic representation. As demonstrated by the lines, the loss exhibits an approximately quadratic increase with the field. Figure 4(c) demonstrates that, in contrast to the high-frequency flank of the $\alpha$ peak, in the EW region there is no significant field dependence. Similar results were also found for other temperatures and for PC.

All these findings demonstrate that the loss in the EW region does not show any nonlinear behavior. What could be the reason for this qualitatively different behavior compared to the right $\alpha$-peak flank? A positive nonlinear effect was first observed in [41] and ascribed to the breaking of correlations between molecules at high fields, leading to a larger effective dipolar moment. Interestingly, recent theoretical considerations also suggest a close connection of cooperativity of relaxational motions and nonlinearity of the dielectric response [42,43]. Within this framework, it is reasonable that nonlinearity is absent in the EW region: secondary relaxations usually are considered as non-cooperative processes, which is corroborated by the Arrhenius temperature-dependence of their relaxation time [32,44,45] (however, see, e.g., [46] for a different view). In contrast, the $\alpha$ relaxation shows strong deviations from Arrhenius behavior, which can be formally described by an increasing energy barrier at low temperatures, commonly ascribed to increasing cooperativity [45,47]. This notion is consistent with the rising peak amplitude in $\Delta \ln \varepsilon''$ with decreasing temperature, found in the present work (Fig. 2 and inset of Fig. 3), signifying increasing nonlinearity in the $\alpha$-peak region when the glass temperature is approached [10].

Finally, one may ask how the present results can be interpreted within the picture developed by Richert and Weinstein for the explanation of the strong nonlinearity of the $\alpha$ relaxation [8]. Their model is based on heterogeneity and, thus, one may speculate that the found absence of nonlinearity could imply the absence of heterogeneity in the EW region. However, this seems unlikely as the wing is caused by a broad relaxation peak [21], indicative of a broad distribution of relaxation times. In addition, in a homogeneous scenario a shift of the whole relaxation peak could be expected. Moreover, DHB experiments have proven that heterogeneity still persists in the region of the $\beta$ relaxation [25]. Interestingly, in [26] a qualitatively different DHB response of $\alpha$ peak and EW was revealed: At high frequencies the persistence time of the burned holes is determined by the burn frequency while close to the $\alpha$ peak it is equal to the structural relaxation time. However, it should be noted that in [26] the EW in dielectric modulus spectra was considered. Thus, it is unclear if these findings, measured at relatively low frequencies, are of relevance for the present permittivity results, where the EW occurs at significantly higher frequencies.

In this context it is interesting that physical aging data on glycerol in the EW region [34] can be described by a homogeneous picture while the aging at the $\alpha$ peak shows heterogeneous characteristics [35]. This was explained assuming that the domains with very short relaxation times, which generate the EW and make up only few percent of the total sample, are completely surrounded by slower-mode domains, solid on the time scale of the EW modes. Thus, the "inner clock" of the fast modes is linked to the macroscopic softening, i.e., the structural relaxation [35]. Within this framework, the missing nonlinearity of the EW found in the present work seems to indicate that these fast modes are not able to take up significant amounts of field energy, which would lead to a decrease of their relaxation times.

In summary, we have found a considerable difference in the nonlinear dielectric properties of the $\alpha$ and EW relaxation of two prototypical glass formers. The found very strong nonlinearity at the high-frequency flank of the $\alpha$ peak is fully consistent with the heterogeneity of relaxation, which was shown to be an inherent property of the glassy state of matter. However, the absence of any nonlinear effect in the EW region implies a qualitative difference of the modes contributing in this region. Either they are not able to absorb significant amounts of field energy or they dissipate this energy to the phonon bath much faster than the local relaxation time governing their dielectric response (however, the latter may be incompatible with the modeling of aging results in ref. [35]). The relation of this unexpected finding to other unusual properties of the EW as the homogeneous aging behavior is unclear. Overall, it remains to be clarified in future work how the heterogeneity-based model can account for the found lack of nonlinearity in the EW region. The present findings, however, are consistent with a recent theory relating nonlinearity and cooperativity of molecular motion [42], which would imply a non-cooperative nature of the $\beta$ relaxation.

This work was supported by the Deutsche Forschungsgemeinschaft via Research Unit FOR1394.

# Supplementary Information for

# Nonlinear dielectric response at the excess wing of glass-forming liquids

Th. Bauer, P. Lunkenheimer,[*] S. Kastner, and A. Loidl

*Experimental Physics V, Center for Electronic Correlations and Magnetism, University of Augsburg, 86135 Augsburg, Germany*

## 1. Experimental details

The sample materials (glycerol of ≥ 99.5 % purity and propylene carbonate of 99.7 % purity, both purchased from Sigma-Aldrich) were arranged between two polished and lapped stainless steel plates. Instead of using spacers, the upper plate (6 mm diameter) floats on the sample material and the lower plate (10 mm diameter). Thus a minimum sample thickness down to 1.1 µm is achieved, which is deduced from a comparison of the absolute values of $\varepsilon''$ with the published results from broadband dielectric spectroscopy [1,2]. In addition, we found that with this sample preparation the probability for electrical breakthroughs was significantly reduced, most likely due to the missing boundaries between sample and spacers.

For the measurements, we used an "Alpha-A" frequency-response analyzer in combination with a high voltage booster "HVB 300", both from Novocontrol Technologies. This setup enables the application of ac signals with peak voltages up to 150 V. Due to the small sample thickness, fields up to about 700 kV/cm are reached. Following the procedure used by Weinstein and Richert [3,4], at each frequency we have performed successive high- and low-field measurements, which were separated by a waiting time. To minimize effects from phonon heating, few high-field oscillations were applied, followed by a cooling period achieved by applying a series of "waiting-oscillations" with 0.01 V only. Subsequently, a low-field measurement with 2.18 V (14 kV/cm) was carried out. At temperatures above 204 K and for fields beyond 500 kV/cm the number of applied "waiting-oscillations" was 46 times higher than the cycle number of the high-field measurement to ensure that the low-field data are not affected by the preceding high-field measurement. In all other measurements the factor was 23.

We found considerable phonon heating when applying too many field cycles, especially at high temperatures and fields (considerably exceeding those in [8,4]). Thus, we reduced the cycle number and used two cycles only at frequencies ν < 2 Hz while at higher frequencies more cycles were applied (there the minimum cycle number is limited by the processing speed of the experimental setup). However, we made sure that phonon heating was excluded in all cases by carefully checking the influence of the cycle number. As noted in [8], trivial phonon heating can also be excluded, based on the fact that an increase of phonon temperature should simply lead to a shift of the entire spectra to higher frequencies, which is in contrast to the actual results (Figs. 1 and 3 of the main article).

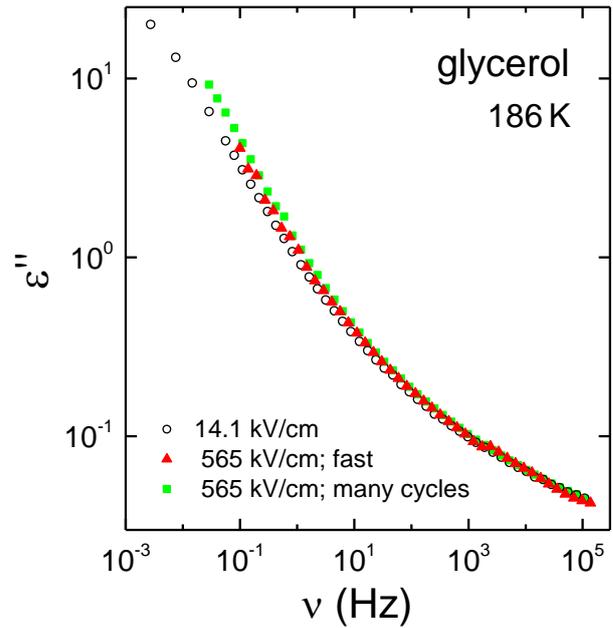

FIG. S1. Frequency dependence of the dielectric loss of glycerol at 186 K. In addition to a low-field measurement (circles), two high-field curves obtained with different cycle numbers are shown (crosses; see text). In the excess-wing region, no significant nonlinear effect is detected, independent of the applied number of field cycles.

When applying a small number of cycles only, the measured dielectric loss may no longer represent the equilibrium dielectric response. In dielectric hole burning experiments, Schiener et al. [5] have demonstrated that at a burn frequency of 1 Hz, about 80% of the equilibrium value is reached after a single oscillation. However, for 20 Hz this number has already increased to ten oscillations. To check if the found absence of nonlinearity in the excess-wing region is a non-equilibrium effect, we performed two high-field measurements with different numbers of field cycles (Fig. S1). One of these measurements was performed in the fastest measurement mode that is possible with our experimental setup (using a faster computer, using an Alpha analyzer with fast data rate option, disabling reference measurements of the Alpha analyzer, etc.). The other measurement was done without taking these measures and explicitly choosing large measurement times in the employed commercial measurement program (Novocontrol Windeta). Typical numbers of cycles (monitored by an oscilloscope) in the latter measurement were about 10 cycles for $\nu < 1$ Hz, 50 cycles for 100 Hz, 2000 cycles for 1 kHz, and $\geq 2\times 10^5$ cycles for 100 kHz. Extrapolating the results of ref. [5], it becomes clear that for these large numbers of cycles, at high frequencies equilibrium must be reached. As revealed by Fig. S1, just as for the fast measurement (triangles), no field-dependent change in the excess-wing region is found for this measurement with many cycles (squares). This clearly proves that this absence of nonlinearity is not due to a non-equilibrium effect.

In contrast to the behavior in the wing region, at the high-frequency flank of the $\alpha$ peak, a somewhat smaller enhancement of $\varepsilon''$ is found for the fast high-field measurement. For example, at 1 Hz only a single cycle was applied in this measurement, which according to [5] is not sufficient to completely reach equilibrium. This explains the somewhat smaller loss compared to the slow measurement, where about 15 oscillations were applied.

For cooling, a closed cycle refrigeration system (CTI-Cryogenics) was used.

## 2. Sample contraction

Applying high voltages can lead to significant attracting forces between capacitor plates and may be suspected to lead to spurious effects in nonlinear dielectric measurements [4,6,7]. With maximum values for the field, $E_{max} = 671$ kV/cm, and for the dielectric constant, $\varepsilon'_{max} = 80$ [39], and using $F_{max} = \varepsilon_0 \varepsilon'_{max} A E_{max}^2/2$ [4], we find a maximum force $F_{max} = 45$ N and a maximum stress $\sigma_{max} = F_{max}/A = 1.6\times 10^6$ Pa. The relative contraction $\Delta d/d$ can be estimated by $\Delta d/d = \sigma_{max}/(3G')$ where the approximation $Y' \approx 3 G'$ was used [8] with $G'$ the shear modulus and $Y'$ denoting Young's modulus. Hence the contraction of the glycerol sample is less than 0.5 % with a worst case estimate of $G' = 1\times 10^8$ Pa according to [9].

To further exclude a contraction effect, we performed additional measurements using glass-fiber spacers with a diameter of 30 μm. To achieve sufficiently high fields, these measurements were performed with both a "HVB 4000" high voltage booster from Novocontrol Technologies and an "aixACCT TF2000" ferroelectric analyzer, applying voltages up to 1.1 kV. Measurements with the ferroelectric analyzer yield time-dependent polarization signals in response to the sinusoidal exiting field. A Fourier analysis of these signals and a comparison with the time-dependent electrical field allows for the calculation of the dielectric permittivity. The results obtained with both devices fully corroborate the finding of the absence of nonlinear effects in the excess-wing region, reported in the main article.

## 3. Excess-wing nonlinearity in the heterogeneity model

Within the heterogeneity model discussed, e.g., in [8], the nonlinearity depends on the activation energy of the relaxation time and on the slope $\partial \log \varepsilon'' / \partial \log \nu$ in the double-logarithmic spectra. Thus, the question may arise if different values of these quantities for the excess wing may trivially explain the approach of zero nonlinearity at high frequencies. As noted in ref. [8], the activation energy determines the shift of the relaxation time (and, thus, of the $\alpha$-peak frequency) caused by the field-induced excess temperature $T_e$. The mentioned slope then determines how much the loss varies due to this shift. The effect of the slope has already been discussed in the explanation of the results on propylene carbonate in ref. [10]. In that work it was used to rationalize the observed onset of a reduction of $\Delta \ln \varepsilon''(\nu)$ at high frequencies. However, for any non-zero slope, this effect cannot lead to $\Delta \ln \varepsilon''(\nu) = 0$ as found in our data that are extending further into the wing region than in ref. [10]. The excess-wing slope is about a factor 2-3 smaller than that of the alpha-relaxation and the resulting smaller $\Delta \ln \varepsilon''(\nu)$ would be well within our detection limit.

How about the influence of the activation energy? The excess wing arises from a secondary relaxation peak, partly superimposed by the $\alpha$ peak [11,12]. It is well known that the energy barriers of secondary relaxations are significantly smaller than those of the $\alpha$ relaxation [13,14,15]. Therefore, at first glance one could assume that the field-induced temperature shift $T_e$ may cause a much smaller horizontal shift of the loss spectra in the wing region and, thus, lead to a smaller or even negligible shift of $\varepsilon''$. However, one should note that the energy barriers reported for secondary relaxations are very often based on data measured below the glass temperature $T_g$, i.e., in non-equilibrium, while the experiments of the present manuscript all were performed above $T_g$. It is well known that below $T_g$, energy barriers are much lower than above $T_g$, which for $\beta$ relaxations was explicitly shown, e.g., in [16]. Moreover, it was shown that in glass formers with excess wing (sometimes termed "type A") like glycerol or propylene carbonate, above $T_g$ the relaxation time of the secondary relaxation, giving rise to the excess wing, has similar temperature dependence as the $\alpha$-relaxation (only the absolute values are somewhat smaller). This was explicitly demonstrated for glycerol, e.g., in [17]. It is this nearly parallel evolution of the relaxation-time curves in the Arrhenius plot above $T_g$, which prevents that the excess-wing relaxation becomes visible as shoulder or separate peak in type A systems above $T_g$ [11]. The fact that the energy barriers of $\alpha$ and excess-wing relaxation are nearly identical can also directly be deduced from the finding that the horizontal difference of the spectra of two neighboring temperatures is nearly identical in the alpha-peak and the excess-wing region (cf., e.g., the 186 K and 195 K curves at low fields in Fig. 1(b) of the main paper).

Thus, an energy barrier of similar magnitude as for the $\alpha$ relaxation has to be used to estimate the shift of the relaxation time for a given $T_e$. Together with the reduced slope in the wing region, at high frequencies this will lead to a frequency independent $\Delta \ln \varepsilon''(\nu)$ that is smaller than the observed peak but still of significant magnitude, in marked contrast to the approach of zero observed in the present work.

**4. Low-frequency response**

At low frequencies, $\nu < \nu_p$, according to the heterogeneity model discussed, e.g., in [8,18], $\Delta \ln \varepsilon''$ should become negative. In our data on glycerol (Fig. 2 of the main article), the experimental data for $\nu < \nu_p$ scatter around zero and a clear statement on the validity of the theoretical prediction is difficult. However, for propylene carbonate (inset of Fig. 3), a minimum in $\Delta \ln \varepsilon''(\nu)$ is found, in accord with the theory. However, $\Delta \ln \varepsilon''$ remains positive for all frequencies, in contrast to the theory. Interestingly, similar behavior was also reported for propylene carbonate in ref. 19. At present it is unclear if this finding is pointing to a failure of the theory or if it is due to an experimental artifact. For example, ionic impurities in polar glass formers can lead to small conductivity contributions at the low-frequency flank of the $\alpha$ loss peaks and in [20] significant nonlinear effects in the ionic conductivity of glass formers were reported. Further experiments with higher resolution on extremely purified samples have to be performed to clarify these questions.

____________________